\documentclass[twoside]{ilcws08}
\usepackage[latin1]{inputenc}
\usepackage[dvips]{graphicx,epsfig,color}
\usepackage{wrapfig,rotating}
\usepackage{amssymb,amsmath,array}

\begin{document}
\title{
Test Beam Requirements for the ILC Tracking and Vertex Detectors} 
\author{Ingrid-Maria Gregor$^1$ 
\vspace{.3cm}\\
1- DESY Hamburg, Notkestr. 85, D-22607 Hamburg, Germany \\
}

\maketitle

\begin{abstract}
In this report the test beam requirements for the vertex detector
and the tracking detector for ILC are discussed. It focuses on the 
infrastructure needs of the different subsystems. In the second part
of this summary the ideas about future infrastructure above the immediate
needs are summarised.
\end{abstract}

\section{Introduction} 

Test beam requirements for tracking and vertex at the ILC are continuously
discussed in the community.
After the test beam workshop at Fermilab in 2007 all needs were 
summarised in a document~\cite{Andy}.
Within the EU supported project EUDET~\cite{EUDET} some test beam 
infrastructure is being build, but this will and can not cover the 
world wide needs for position sensitve detectors as for the vertex and 
tracking detectors will be needed. In this report the requirements as 
know before the LCWS08 in Chicago are summarised.

\section{Vertex Detector}
 
\begin{wraptable}{l}{0.37\columnwidth}
\centerline{\begin{tabular}{|l|c|c|}
\hline
Accelerator & a [$\mu$m] &  a [$\mu$m] \\ \hline \hline
LEP	& 25	& 70 \\ \hline
SLD	&8	&33 \\ \hline
LHC	&12	&70 \\ \hline
RHIC-II	&15	&19 \\ \hline
ILC	& $<$5	& $<$10 \\ \hline 
\end{tabular}}
\caption{Impact parameter resolution at different
accelerators.}
\label{tab:fom}
\end{wraptable}
The ILC vertex detector will be build to measure the impact parameter, the
charge for every charged track in jets, and 
the vertex mass. For this a good angular coverage with many layers close to 
vertex is needed ($| \cos{\theta}|< 0.96.$). It is planned to have five
layers distributed between 15~mm and 60~mm radius from the beam axis.
An efficient detector is needed for a very good impact parameter resolution, 
and the material should no exceed
$~0.1\% X_0$ per layer. Above that the sensors should be capable to cope
with the ILC beamstrahlungs background, and the power consumption should be
modest, optimally less than 100 Watts. In order to meet the above requirements a vertex detector
with small pixels, thin sensors, thin readout electronics, low power 
(gas cooling) needs to be build.  A hit resolution of better than 5$\mu$m
is required.
A figure of merit for the ILC vertex detector is the impact parameter
resolution which can be parametrised as
\begin{math}
\sigma_{r\phi}\approx\sigma_{rz} \approx a \oplus \frac{b}{b\sin^{3/2}\theta}  
\end{math}.  
In table~\ref{tab:fom} the parameters $a$ and $b$ for all major HEP 
experiments are listed. This shows that the requirements at the ILC vertex 
are unprecedented. Currently there are about ten candidate sensor technologies
for the ILC vertex detector. A summary about all these approaches is given 
in~\cite{chris}. 

\subsection{Important test towards a technology choice}
One of the next major steps for all possible technologies is building ladders
to show not only the single pixel performance but also the interfacing to
the infrastructure, and the mechanical integration.
At the level of single ladders an number of parameters have to studied at
test beams:  
signal-over-noise, single point resolution, efficiency, double hit separation.
Also the homogeneity of the detection,the read out and the data handling have
to be tested. It also may be of interest to run within "sizeable¨ magnetic 
field in order to assess effects on cluster characteristics (e.g. single point resolution).
In a further step with multiple layers additionally to the above tests the
standalone tracking capabilities, tracking under high occupancy, and
low momentum tracking would needed to be studied.
Possible further tests are studying the homogeneity of the performances over 
the ladder surface, the multi-channel and multi-chip operation, the electrical
servicing of the chip and the cooling system operation. The last issue include
mechanical properties and the influence on the performances such as sagitta 
and vibrations vs. single point resolution. Also the heating versus the 
signal-to-noise ratio or fake hits (noisy pixel rate) should be investigated.

\subsection{Needed Test Beams}
A high energy hadron beam ($\sim$100 GeV) would be needed for position 
resolution testing and a low energy beam for the low momentum tracking.
The beam spots should be adjustable from a about 1~mm$^2$ to a few cm$^2$.
A "ILC-like" spill structure could be useful (1~ms beam at 200~ms intervals) 
to see the effects on the read out when particles arrive and to allow a read 
out during a "quite" phase. Such a beam structure is not needed for all tests
but such a beam would give the opportunity to test all technologies under 
ILC like conditions.  
This might be necessary close to the technology decision.
\begin{wraptable}{l}{0.7\columnwidth}
\centerline
{\begin{tabular}{|l|c|c|c|}
\hline
Lab. 	& $E_{b}$ [GeV] & Particles	& Availability \\ \hline \hline
CERN PS		& 1 - 15	& e, h, m	& LHC  prior. \\  \hline
CERN SPS	&10 - 400	& e, h, m 	& LHC  prior. \\ \hline
Fermilab	& 1-120	& e, p, K; m	& contin. (5\%)\\ \hline
\end{tabular}}
\caption{Test beam facilities, energy range and availabiliy which are useful 
for the necessary vertex studies.}
\label{tab:beams}
\end{wraptable}
Such test beam not available right now and investment would be needed. 
It is technically feasible and Fermilab is currently following up this issue.
In table~\ref{tab:beams} the test beam facilities suitable for the above
tests are mentioned. 

\section{Tracking Detector}
Immediately outside the vertex detector a tracking system will follow.
The options considered for the tracking detector are large silicon trackers 
(à la ATLAS/CMS) and Time Projection Chambers (TPC) with $\sim$100~$\mu$m 
point resolution (complemented by Si-strip devices). 
The performance goals for this systems are:
Continuous 3D tracking, easy pattern recognition throughout volume,
~98-99\% tracking efficiency in presence of backgrounds,
time stamping to 2ns together with inner silicon layer,
minimum of X$_0$ inside ECAL (~3\% barrel, ~15\% endcaps),
$\sigma_{pt}\sim$50~mm (r$\phi$) and $\sim$500~mm (rz) \@ 4T,
two-track resolution $<$2mm (r$\phi$) and $<$5~mm(rz), and
dE/dx resolution $<$5\%.
In the following the test beam infrastructure for both concepts are
summarised separately. 

\subsection{Gaseous Tracker}
The groups working on a gaseous tracker for ILC are currently taking 
advantage of the EUDET infrastructure of a 1~Tesla Magnet at the 6~GeV 
electron beam at DESY. A full field cage is set up in the magnet.
But this system would  after initial tests need to be moved to a 
$\sim$ 50~GeV/c, hadron beam. Preferable a mixed hadron beams with particle 
ID for dE/dx. The intensity should be variable from low to high. As the LCTPC
has a Silicon tracker already included, additional tracking systems are not
urgently needed. A number of studies should be performed in a large volume 
high field magnet ($\sim$ 2~T), but not necessarily at the test beam facility. 
RD51 is planning a dedicated beam area for micro gas detectors at CERN SPS
also want to provide infrastructure to help test of gaseous detectors \\

\subsection{Si Tracker}
The Si-tracker group expressed that the existent test beams are 
adequate for their needs but improvement in the overall infrastructure
is required. A beam telescope and associated DAQ and trigger logic such as
the EUDET pixel telescope would be of advantage. A general DAQ framework in
the ILC fashion would help to accelerate the turn around of test beam results.
A high field magnet with up 3 Teslas would be needed at a later point, but not
necessarily for all test beams.
Furthermore it was expressed that access to a mechanical workshop and 
support for last minute needs during installation would be very helpful.
Further items of use are: 3D table(s) to install and properly move the 
prototypes wrt the beam, easy access to computing facilities and LabNet, 
control room(s) with enough space (for several users), lab staff responsible 
for the good running of the test beam, and a crane to install and move heavy 
prototypes. Most of these items are available at the test beams but generally
not easy to access for users from outside institutes. The ILC community should
help to improve the situation. 

\section{Beyond the immediate needs}
The next logical step for ILC is assess system aspects of the proposed 
detector concepts. The principle integrating factor in linear collider event 
reconstruction is the concept of ¨energy flow¨ where the reconstructed 
objects from different detectors are combined into physics objects such as 
leptons, photons, or jets. The particle flow approach relies on robust 
identification and precise momentum measurement of charged particles. 
It needs to be established how to form these particle-flow objects, 
how to do the mechanical  integration, and also the common DAQ systm.
This requires the definition of interfaces and their implementation. 
In the beginning of 2007 a group of European research institutes involved
in ILC detector R\&D submitted a proposal to the EU to work on this 
concept. The
idea of EUVIF idea was to build an unique infrastructure to integrate 
commensurate prototypes of LC detector components and install this in a 
test beam with different particles and in appropriate energy range. The 
proposal was not granted, but within the proposal the ideas for such a concept
where phrased and describes essential what is needed by the vertex and 
tracking groups beyond the immediate needs described above. 


\subsection{Vertex}
A small-scale full vertex detector could be build. interface 
to be able to replace the ladders by different type ladders
Building a global mechanical infrastructure to host multi-layer modules for vertex detectors in different technologies (design independent)
Developing the data acquisition system including hardware from EUDET to suit the new infrastructure
Software for reconstruction and analysis (calibration, alignment, pattern recognition) -> based on existing software for EUDET telescope
Producing a target system to create jet-like structures
Integrating the EUDET telescope upstream of the target

\subsection{Tracker}
\subsubsection{Intermediate Tracker infrastructure}
Lightweight structures for both module carrier and overall support
Prototype silicon small area modules equipped with single sensors up to daisy-chained ladders 
Overall support structure for modules/ladders arranged in layers (leightweigth, ultra-thin)
Improving the existing EUDET readout chip and developing a front-end hybrid prototype suitable for testing silicon sensors with conventional (wire-bonding) or novel (bump-bonding) techniques
Integration of the front-end electronics developed in EUDET into the central DAQ system

\subsubsection{Gaseous Infrastructure}
The EUDET TPC infrastructures could be provided for combined tests of the 
particle flow concept. This would allow the optimisation of the overall 
detector design. An interface to the common DAQ and slow control system 
should be faciliated. 
Within EUVIF is was planned to develop and provide the readout software,
to improve a slow control system, and to integrate in th overall EUVIF system.

\section{Summary}
In this report the immediate needs of the ILC tracking and vertex community
for test beam infrastrucutre is summarised. In the second part of the report
the ideas of the so-called EUVIF proposal are described. This proposal
was forseen to provide test beam infrastructure above the immediate needs.
The proposal was not granted but other funding agencies might 
accept it in the future.

\section{Bibliography} 


\begin{footnotesize}



%

\end{footnotesize}


\end{document}